\documentclass[a4paper]{article}

\usepackage{INTERSPEECH2019}
\usepackage{amsmath,graphicx,indentfirst,url,dirtree}
\usepackage{CJKutf8}

\newcommand{\Fig}[1]{Fig. \ref{fig:#1}} % refer to figure
\newcommand{\Table}[1]{Table \ref{tb:#1}} % refer to table
\newcommand{\Figure}[3]{\vspace{-0mm} \includegraphics[width=#1,clip]{#2} \vspace{-0mm} \caption{#3} \vspace{-0mm} \label{fig:#2}}

\newcommand{\drawfig}[4]{ % draw figure 
  \begin{figure}[#1]
  \begin{center}
  \Figure{#2}{#3}{#4}
  \end{center} 
  \end{figure}
}
\title{JVS corpus: free Japanese multi-speaker voice corpus}

\name{Shinnosuke Takamichi, Kentaro Mitsui, Yuki Saito, \\Tomoki Koriyama, Naoko Tanji, and Hiroshi Saruwatari}
\address{Graduate School of Information Science and Technology, The University of Tokyo, \\
7-3-1 Hongo Bunkyo-ku, Tokyo 133--8656, Japan.}

\email{shinnosuke\_takamichi@ipc.i.u-tokyo.ac.jp}

\begin{document}
\ninept
\maketitle
\begin{abstract}
Thanks to improvements in machine learning techniques, including deep learning, speech synthesis is becoming a machine learning task. To accelerate speech synthesis research, we are developing Japanese voice corpora reasonably accessible from not only academic institutions but also commercial companies. In 2017, we released the JSUT corpus, which contains 10 hours of reading-style speech uttered by a single speaker, for end-to-end text-to-speech synthesis. For more general use in speech synthesis research, e.g., voice conversion and multi-speaker modeling, in this paper, we construct the JVS corpus, which contains voice data of $100$ speakers in three styles (normal, whisper, and falsetto). The corpus contains 30 hours of voice data including 22 hours of parallel normal voices. This paper describes how we designed the corpus and summarizes the specifications. The corpus is available at our project page\footnote{\url{https://sites.google.com/site/shinnosuketakamichi/research-topics/jvs_corpus}}.  
\end{abstract}


\section{Introduction}
Thanks to developments in deep learning techniques, studies on speech have been targeted actively \cite{hinton12dnnasr,oord16wavenet,takamichi17moment,saito18advss}. Nowadays, speech synthesis, e.g., text-to-speech, singing voice synthesis, voice conversion, and speech coding, is becoming a machine learning task. Easily accessible voice corpora help to not only accelerate speech-related research but also improve the reproductivity of a study. In 2017, we released a large-scaled Japanese speech corpus, named the JSUT corpus \cite{sonobe17jsut}, for end-to-end text-to-speech synthesis. The corpus included 10 hours of reading-style speech data uttered by a single native Japanese speaker and all pronunciations of daily-use characters and individual readings in Japanese \cite{joyokanji}. Since Oct. 2017, the project page \cite{jsut_corpus} was accessed more than 6,000 times (75\% from Japan and 25\% from foreign countries) from more than 60 countries. We believe that the JSUT corpus has become one of the most used Japanese corpora for modern speech synthesis research \cite{ueno19multispeakerend2endtts,luo19waveletf0feature}.

Towards more general purposes of speech-related research, this paper introduces a new Japanese voice corpus, named the \textit{JVS} (Japanese versatile speech) corpus. The corpus is designed to have many benefits for many types of users as follows.
    \begin{itemize} \leftskip -5mm \setlength{\itemsep}{-0pt} 
    \item []\textbf{High-quality format}: The audio files are sampled at 24~kHz, encoded at 16~bit, and formatted in RIFF WAV.
    \item []\textbf{High-quality recording}: The recordings were controlled by a professional sound director and done in a recording studio. 
    \item []\textbf{Many speakers}: The corpus includes 100 native Japanese speakers, and all of the speakers are professional, e.g., voice actor/actress.
    \item []\textbf{Many styles}: Each speaker utters not only normal speech but also whisper and falsetto voices.
    \item []\textbf{Large in scale}: In total, the corpus contains 30 hours of voice data.
    \item []\textbf{Parallel/non-parallel utterances}: Each speaker utters parallel, i.e., common among speakers, and non-parallel, i.e., completely different among speakers, utterances.
    \item []\textbf{Many tags}: The corpus includes not only voice data but also transcriptions, gender information, $F_0$ ranges, speaker similarity, and phoneme alignments.
    \item []\textbf{Free for research}: The corpus is free to use for research in academic institutions and commercial companies.
    \item []\textbf{Easily accessible}: The corpus is freely downloadable online.
    \end{itemize}
The next section describes how we designed the corpus.

\section{Corpus design}
The corpus consists of the following four sub-corpora. Their names are formatted as \textit{[NAME][NUM\_UTT]}. \textit{[NUM\_UTT]} indicates the number of utterances per speaker.
    \begin{itemize} \leftskip -5mm \setlength{\itemsep}{-1pt}
    \item[] \textbf{parallel100}: 100 parallel normal (reading-style) utterances
    \item[] \textbf{nonpara30}: 30 non-parallel normal utterances
    \item[] \textbf{whisper10}: 10 whisper utterances
    \item[] \textbf{Falsetto10}: 10 falsetto utterances
    \end{itemize}
The directory structures of the corpus are listed below. The speaker name is formatted as \textit{jvs[SPKR\_ID]}. \textit{[SPKR\_ID]} indicates the speaker ID with the range of 1 through 100.   
    \dirtree{% 
    .1 \includegraphics[width=0.25cm]{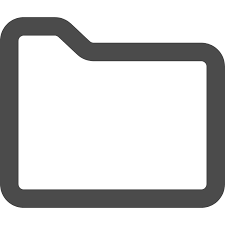} \textbf{jvs001}.
        .2 \includegraphics[width=0.25cm]{eps/dir.png} \textbf{parallel100}.
            .3 \includegraphics[width=0.25cm]{eps/dir.png} wav24kHz16bit. 
            .3 \includegraphics[width=0.25cm]{eps/dir.png} lab.
            .3 \includegraphics[width=0.25cm]{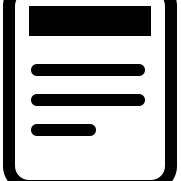} transcripts\_utf8.txt.
        .2 \includegraphics[width=0.25cm]{eps/dir.png} \textbf{nonpara30}.
            .3 \includegraphics[width=0.25cm]{eps/dir.png} wav24kHz16bit. 
            .3 \includegraphics[width=0.25cm]{eps/dir.png} lab.
            .3 \includegraphics[width=0.25cm]{eps/file.png} transcripts\_utf8.txt.
        .2 \includegraphics[width=0.25cm]{eps/dir.png} \textbf{whisper10}.
            .3 \includegraphics[width=0.25cm]{eps/dir.png} wav24kHz16bit.
            .3 \includegraphics[width=0.25cm]{eps/file.png} transcripts\_utf8.txt.
        .2 \includegraphics[width=0.25cm]{eps/dir.png} \textbf{falsetto10}.
            .3 \includegraphics[width=0.25cm]{eps/dir.png} wav24kHz16bit. 
            .3 \includegraphics[width=0.25cm]{eps/file.png} transcripts\_utf8.txt.
    .1 \includegraphics[width=0.25cm]{eps/dir.png} \textbf{jvs002}.
    .1 \includegraphics[width=0.25cm]{eps/dir.png} ....
    .1 \includegraphics[width=0.25cm]{eps/dir.png} \textbf{jvs100}.
    .1 \includegraphics[width=0.25cm]{eps/file.png} speaker\_similarity\_male.csv. % 話者類似度行列 (齋藤SSW)
    .1 \includegraphics[width=0.25cm]{eps/file.png} speaker\_similarity\_female.csv. % 話者類似度行列 (齋藤SSW)
    .1 \includegraphics[width=0.25cm]{eps/file.png} duration.txt. % 話者毎の音声データ量
    .1 \includegraphics[width=0.25cm]{eps/file.png} gender\_f0range.txt. % 話者毎のF0上限・下限
    }

    \subsection{Sub-corpora}
        This section describes how we designed the four sub-corpora. 
	
	    \subsubsection{parallel100}
	    Parallel voices, i.e., utterances that are common among speakers, are used for voice conversion \cite{toda07_MLVC,stylianou88}, speaker factorization \cite{lu13factor}, multi-speaker modeling \cite{ueno19multispeakerend2endtts}, and so on. We used 100 phonetically-balanced sentences of the sub-corpus ``voiceactress100''\footnote{The original sentences are included in the Voice Actress Corpus \cite{voiceactresscorpus}, and the one included in the JSUT corpus had commas added at the phrase break positions.} of the JSUT corpus \cite{sonobe17jsut}, and we let speakers utter the sentences. This corpus contains not only the audio files but also the transcriptions (stored in ``parallel100/transcript\_utf8.txt'') and phoneme alignment (stored in ``parallel100/lab'').
	    
	    \subsubsection{nonpara30}
	    The use of non-parallel voices, i.e., utterances that are completely different among speakers, is a challenging but more realistic situation than that of parallel voices. Sentences to be uttered are randomly selected from the JSUT corpus excluding its sub-corpus ``voiceactress100.'' Each speaker uttered 30 utterances that are different among speakers. This sub-corpus also includes transcriptions and phoneme alignments. Note that, the sentences are not phonetically balanced unlike the sub-corpora ``parallel100.'' 
	    
	    \subsubsection{whisper10}
	    Whispering is used to quietly communicate, i.e., convey secret information without being overheard. Analysis \cite{ito05whisperanalysis}, synthesis \cite{petrushin10whispertts}, recognition \cite{jou05whisperrecognition} and conversion \cite{toda12bodyconductedvc} of whispered voices have the potential to augment our silent-speech communication. The first five sentences of this sub-corpus are the same as those of the sub-corpus ``parallel100,'' and they are parallel among speakers. The remaining five sentences are the same to those of the sub-corpus ``nonpara30,'' and they are non-parallel among speakers. Namely, ten utterances per speaker are parallel between whispered voices and normal voices.
	    
	    \subsubsection{Falsetto10}
	    Falsetto is a vocal register occupying the $F_0$ range that is higher than normal voices. The physiology of falsetto is different from that of normal voices \cite{childers91vocalqualityfactor}, and the analysis and synthesis of falsetto are remaining tasks for signal processing-based vocoders. The first five sentences of this sub-corpus are the same as those of the sub-corpus ``parallel100.'' The remaining five sentences are the same as those of the sub-corpus ``nonpara30'' but different to those of the sub-corpus ``whisper10.'' Namely, five utterances are parallel among speakers, ten are parallel between normal voice and falsetto, and five are parallel between whisper and falsetto. 
	    
    \subsection{Tags}
        This section describes some of the annotation results. 
        
        \begin{itemize}
            \item \textbf{$F_0$ range (gender\_f0range.txt)}: Typical pitch extractors, e.g., \cite{kawahara99,morise16world,reaper}, have a range for $F_0$ search, and the setting is critical for the results ultimately obtained for the voices. This corpus contains manually annotated $F_0$ ranges per speaker for his/her normal voices.
            \item \textbf{Speaker similarity (speaker\_similarity\_*.csv)}: Perceptual similarity between speakers is useful for selecting speakers (or models) \cite{lanchantin14mavm} and modeling speaker space \cite{saito19perceptual}. This corpus contains perceptual similarity scores between all pairs of speakers of each gender. 
            \item \textbf{Duration (duration.txt)}:  Duration, i.e., data size, and speech rate are also included. Phoneme-level duration is calculated from the results of phoneme alignments.
        \end{itemize}
        
\section{Results of data collection}	
            \begin{table*}[t]
            \centering
            \caption{Speaker-wise duration statistics. Silence parts were included to calculate these values.}
            \label{tb:duration_stats}
            \begin{tabular}{|c||c|c|c|c|c|}
            \hline
                                        & Mininum [min.]         & Average [min.]         & Maximum [min.]         & Total (100 speaker) [hour]      \\ \hline\hline
            parallel100 (100 utterances)                 & 10.11 (jvs020)     &  13.11  & 18.24  (jvs084)    & 22           \\ \hline
            nonpara30  (30 utterances)                  & 2.12  (jvs099)     &  2.62  & 3.86  (jvs036)     & 4.4            \\ \hline
            whisper10  (10 utterances)                  & 0.95 (jvs045)      &  1.24  & 1.69  (jvs018)     & 2.0            \\ \hline
            falsetto10  (10 utterances)                 & 0.90 (jvs045)      &  1.18  & 1.61  (jvs035)     & 2.0            \\ \hline\hline
            \multicolumn{1}{|c||}{Total} & \multicolumn{1}{c|}{-} & \multicolumn{1}{c|}{-} & \multicolumn{1}{c|}{-} & 30.4          \\ \hline
            \end{tabular}
            \end{table*}

	\subsection{Corpus specs}
    We hired 100 native Japanese professional speakers, which included 49 male and 51 female speakers. Their voices were recorded in a recording studio. Recording for each speaker was done within one day. The recordings were controlled by a professional sound director. The voices were originally sampled at 48~kHz and downsampled to 24~kHz by SPTK \cite{sptk}. The 16-bit/sample RIFF WAV format was used. Sentences (transcriptions) were encoded in UTF-8. The full context and monophone labels were automatically generated by Open JTalk \cite{ojtalk}. The phoneme alignments were automatically generated by Julius \cite{lee01julius}. $F_0$ ranges were manually annotated in accordance with hands-on voice conversion \cite{toda19vchandson}. The WORLD vocoder \cite{morise16world,morise16d4c} extracted $F_0$. Commas were added between breath groups. For annotating perceptual similarity scores, we followed Saito et al.'s study \cite{saito19perceptual} and used a crowdsourcing service, Lancers \cite{lancers}, which is a famous crowdsourcing service in Japan. Each listener scored the perceptual similarity for each pair of speakers from $-3$ (completely different) and $+3$ (very similar). A final score for each speaker pair was obtained by averaging listeners' scores. Ten different listeners scored each speaker pair, and 1,000 listeners participated in total. 

	\subsection{Analysis}
	    \subsubsection{Duration}
        \Table{duration_stats} lists the statistics for speaker-wise duration. This corpus contains 26 hours of normal voices and 4 hours of other-style voices. Each speaker uttered approximately 15.7 minutes of normal voices, 1.24 minutes of whispered voices, and 1.18 minutes of falsetto. In the sub-corpus ``parallel100,'' the transcription was common among speakers, but the duration was very different; speaker ``jvs084'' uttered 1.8 times slower than speaker ``jvs020.''

	    \subsubsection{Perceptual speaker similarity}
        \Fig{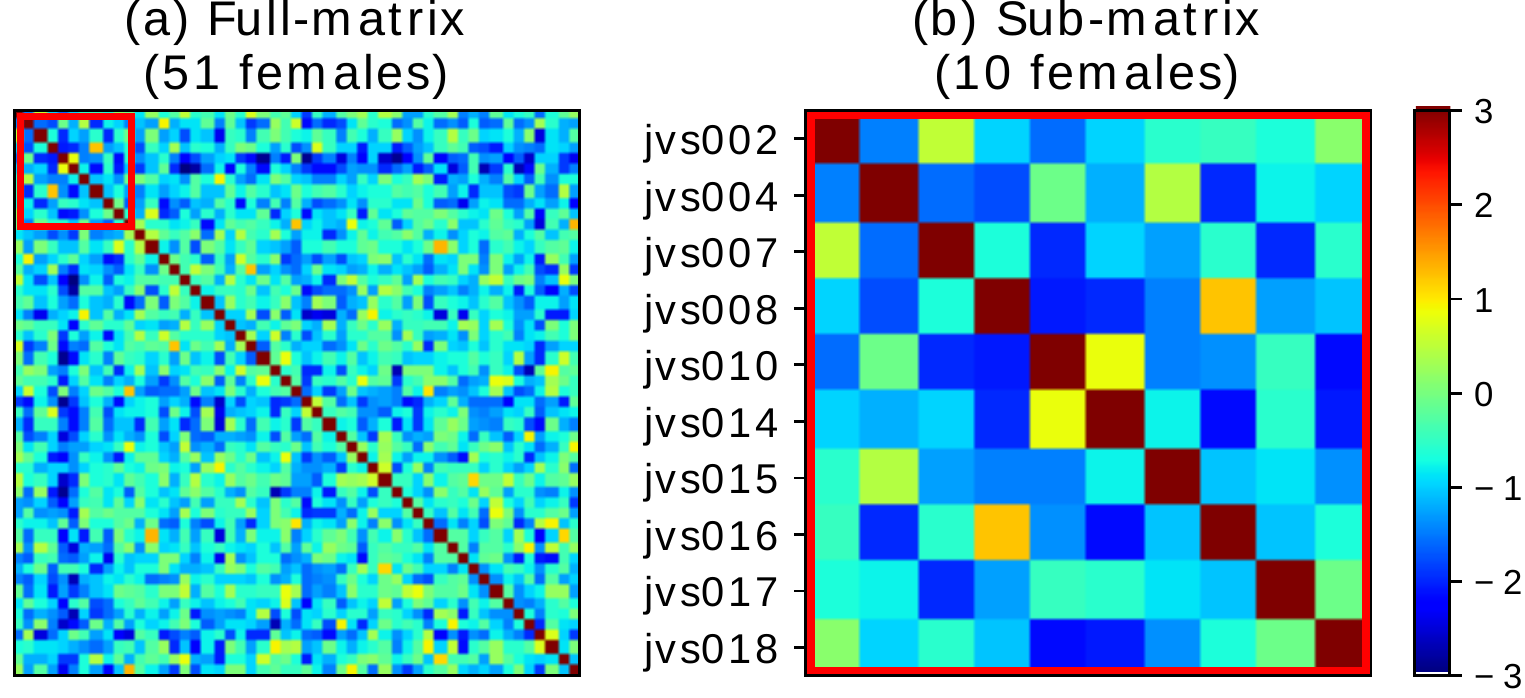} shows matrices of perceptual similarity scores. For example, the most similar pair was ``jvs019'' and ``jvs096.'' Also, a speaker that was most dissimilar from the other speakers was ``jvs010.'' 

\section{Conclusion}	
In this paper, we constructed a corpus named the JVS corpus. The corpus was designed for speech-related research using multi-speaker and multi-style voices. Text data of the corpus is licensed as shown in the LICENCE file in the JSUT corpus \cite{jsut_corpus}. The tags are licensed with CC BY-SA 4.0. The audio data may be used for 
\begin{itemize} \itemsep 0mm
    \item Research by academic institutions
    \item Non-commercial research, including research conducted within commercial organizations
    \item Personal use, including blog posts.
\end{itemize}
Our project page at \url{https://sites.google.com/site/shinnosuketakamichi/research-topics/jvs_corpus} describes the terms for commercial use.

% acknowledgement
\section{Acknowledgements}
Part of this work was supported by the GAP foundation program of the University of Tokyo and the MIC/SCOPE \#182103104.

        \drawfig{t}{1.0\linewidth}{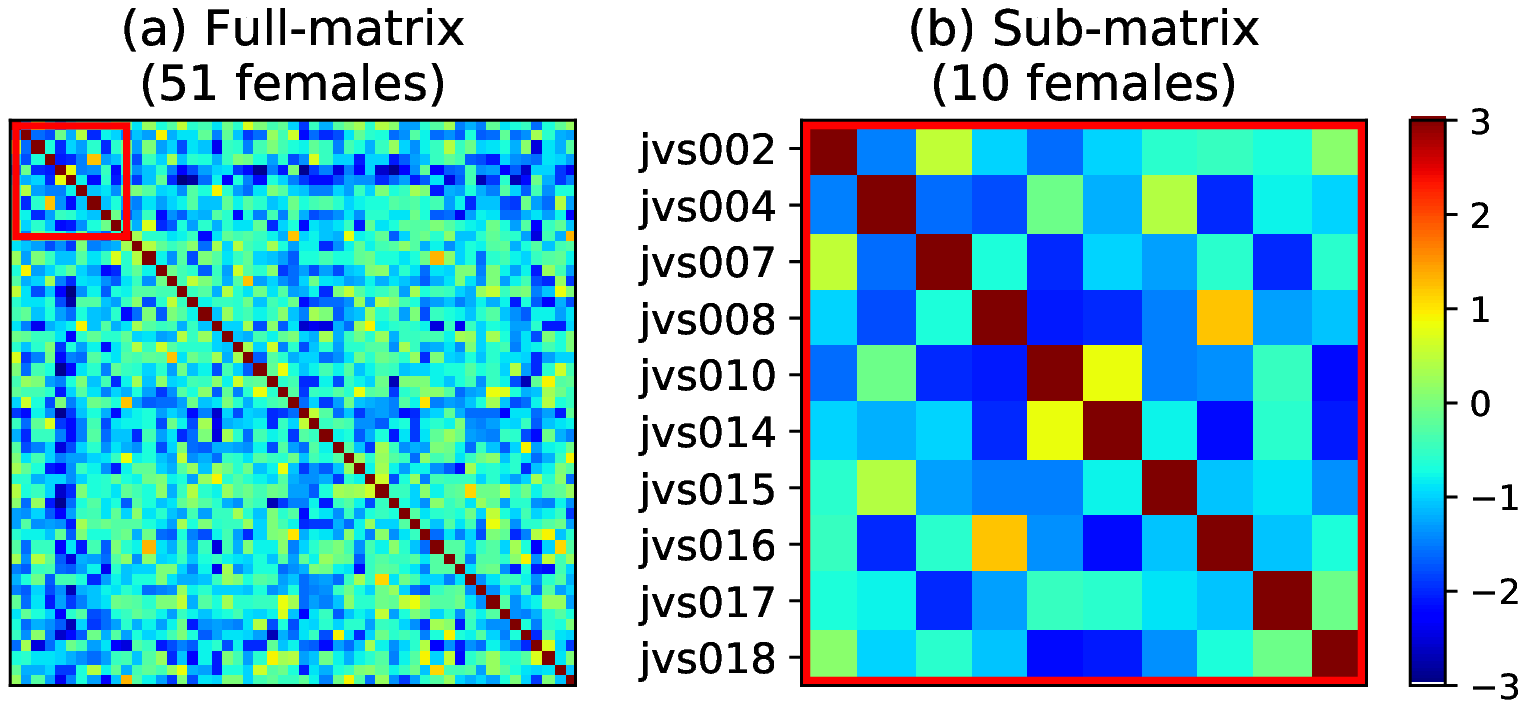}
        {Speaker similarity matrix of 51 Japanese females and (b) its sub-matrix obtained by large-scale subjective scoring.}

\ninept
\bibliographystyle{IEEEbib}
\bibliography{tts}

\begin{thebibliography}{10}

\bibitem{hinton12dnnasr}
G.~Hinton, L.~Deng, D.~Yu, G.~Dahl, A.~r.~Mohamed, N.~Jaitly, A.~Senior,
  V.~Vanhoucke, P.~Nguyen, T.~Sainath, and B.~Kingsbury,
\newblock ``Deep neural networks for acoustic modeling in speech recognition:
  The shared views of four research groups,''
\newblock {\em Signal Processing Magazine of IEEE}, vol. 29, no. 6, pp. 82--97,
  2012.

\bibitem{oord16wavenet}
A.~v.~d. Oord, S.~Dieleman, H.~Zen, K.~Simonyan, O.~Vinyals, A.~Graves,
  N.~Kalchbrenner, A.~W. Senior, and K.~Kavukcuoglu,
\newblock ``Wave{N}et: {A} generative model for raw audio,''
\newblock vol. abs/1609.03499, 2016.

\bibitem{takamichi17moment}
S.~Takamichi, K.~Tomoki, and H.~Saruwatari,
\newblock ``Sampling-based speech parameter generation using moment-matching
  network,''
\newblock in {\em Proc. INTERSPEECH}, Stockholm, Sweden, Aug. 2017.

\bibitem{saito18advss}
Y.~Saito, S.~Takamichi, and H.~Saruwatari,
\newblock ``Statistical parametric speech synthesis incorporating generative
  adversarial networks,''
\newblock {\em IEEE/ACM Transactions on Audio, Speech, and Language
  Processing}, vol. 26, no. 1, pp. 755--767, Jun. 2018.

\bibitem{sonobe17jsut}
R.~Sonobe, S.~Takamichi, and H.~Saruwatari,
\newblock ``{JSUT} corpus: free large-scale japanese speech corpus for
  end-to-end speech synthesis,''
\newblock vol. arXiv abs/1711.00354, 2017.

\bibitem{joyokanji}
Governments of~Japan Agency~for Cultural~Affairs,
\newblock ``List of daily-use kanjis
  \url{http://www.bunka.go.jp/kokugo_nihongo/sisaku/joho/joho/kijun/naikaku/kanji/index.html},''
\newblock 2010.

\bibitem{jsut_corpus}
``{JSUT}: {J}apanese speech corpus of {S}aruwatari {L}ab, the {U}niversity of
  {T}okyo corpus,''
  \url{https://sites.google.com/site/shinnosuketakamichi/publication/jsut}.

\bibitem{ueno19multispeakerend2endtts}
S.~Ueno, M.~Mimura, S.~Sakai, and Tatsuya Kawahara,
\newblock ``Multi-speaker sequence-to-sequence speech synthesis for data
  augmentation in acoustic-to-word speech recognition,''
\newblock in {\em Proc. ICASSP}, Brighton, United Kingdom, May 2019, pp.
  6161--6165.

\bibitem{luo19waveletf0feature}
Z.~Luo, J.~Chen, T.~Takiguchi, and Y.~Ariki,
\newblock ``Emotional voice conversion using dual supervised adversarial
  networks with continuous wavelet transform {F0} features,''
\newblock {\em IEEE/ACM Transactions on Audio, Speech, and Language
  Processing}, vol. 27, no. 10, pp. 1535--1548, Oct. 2019.

\bibitem{toda07_MLVC}
T.~Toda, A.~W. Black, and K.~Tokuda,
\newblock ``Voice conversion based on maximum likelihood estimation of spectral
  parameter trajectory,''
\newblock {\em IEEE Transactions on Audio, Speech, and Language Processing},
  vol. 15, no. 8, pp. 2222--2235, 2007.

\bibitem{stylianou88}
Y.~Stylianou, O.~Capp\'{e}, and E.~Moulines,
\newblock ``Continuous probabilistic transform for voice conversion,''
\newblock {\em IEEE Transactions on Speech and Audio Processing}, vol. 6, no.
  2, pp. 131--142, Mar. 1998.

\bibitem{lu13factor}
H.~Lu and S.~King,
\newblock ``Factorized context modeling for {T}ext-to-{S}peech synthesis,''
\newblock in {\em Proc. ICASSP}. Vancouver, Canada, May 2013.

\bibitem{voiceactresscorpus}
y\_benjo and MagnesiumRibbon,
\newblock ``Voice-actress corpus,'' \url{http://voice-statistics.github.io/}.

\bibitem{ito05whisperanalysis}
T.~Ito, K.~Takeda, and F.~Itakura,
\newblock ``Analysis and recognition of whispered speech,''
\newblock {\em Speech Communication}, vol. 45, no. 2, pp. 139--152, 2005.

\bibitem{petrushin10whispertts}
V.~A. Petrushin, L.~I. Tsirulnik, and V.~Makarova,
\newblock ``Whispered speech prosody modeling for {TTS} synthesis,''
\newblock Chicago, U.S.A., May 2010.

\bibitem{jou05whisperrecognition}
S.-C. Jou, T.~Schultz, and A.~Waibel,
\newblock ``Whispery speech recognition using adapted articulatory features,''
\newblock in {\em Proc. ICASSP}, Philadelphia, U.S.A., Mar. 2005, vol.~1, pp.
  1009--1012.

\bibitem{toda12bodyconductedvc}
T.~{Toda}, M.~{Nakagiri}, and K.~{Shikano},
\newblock ``Statistical voice conversion techniques for body-conducted unvoiced
  speech enhancement,''
\newblock {\em IEEE Transactions on Audio, Speech, and Language Processing},
  vol. 20, no. 9, pp. 2505--2517, Nov. 2012.

\bibitem{childers91vocalqualityfactor}
D.~G. Childers and C.~K. Lee,
\newblock ``Vocal quality factors: Analysis, synthesis, and perception,''
\newblock {\em The Journal of the Acoustical Society of America}, vol. 90, no.
  5, pp. 2394--2410, Nov. 1991.

\bibitem{kawahara99}
H.~Kawahara, I.~Masuda-Katsuse, and A.~D. Cheveigne,
\newblock ``Restructuring speech representations using a pitch-adaptive
  time-frequency smoothing and an instantaneous-frequency-based {F}0
  extraction: Possible role of a repetitive structure in sounds,''
\newblock {\em Speech Communication}, vol. 27, no. 3--4, pp. 187--207, 1999.

\bibitem{morise16world}
M.~Morise, F.~Yokomori, and K.~Ozawa,
\newblock ``{WORLD}: a vocoder-based high-quality speech synthesis system for
  real-time applications,''
\newblock {\em IEICE transactions on information and systems}, vol. E99-D, no.
  7, pp. 1877--1884, 2016.

\bibitem{reaper}
D.~Talkin,
\newblock ``{REAPER: Robust Epoch And Pitch EstimatoR},''
  \url{https://github.com/google/REAPER}.

\bibitem{lanchantin14mavm}
P.~Lanchantin, Mark~J.F. Gales, S.~King, and J.~Yamagishi,
\newblock ``Multiple-average-voice-based speech synthesis,''
\newblock in {\em Proc. ICASSP}, Florence, Italy, May 2014, pp. 285--289.

\bibitem{saito19perceptual}
Y.~Saito, S.~Takamichi, and H.~Saruwatari,
\newblock ``{DNN}-based speaker embedding using subjective inter-speaker
  similarity for multi-speaker modeling in speech synthesis,''
\newblock in {\em Proc. SSW10}. Vienna, Austria, Sep. 2019.

\bibitem{sptk}
``Speech signal processing toolkit ({SPTK}),''
  \url{http://sp-tk.sourceforge.net/}.

\bibitem{ojtalk}
``Open jtalk,'' \url{http://open-jtalk.sourceforge.net/}.

\bibitem{lee01julius}
A.~Lee, T.~Kawahara, and K.~Shikano,
\newblock ``Julius --- an open source real-time large vocabulary recognition
  engine,''
\newblock in {\em Proc. EUROSPEECH}, Aalborg, Denmark, Sep. 2001, pp.
  1691--1694.

\bibitem{toda19vchandson}
T.~Toda,
\newblock ``Hands on voice conversion,''
  \url{https://www.slideshare.net/NU_I_TODALAB/hands-on-voice-conversion},
  2018.

\bibitem{morise16d4c}
M.~Morise,
\newblock ``{D4C}, a band-aperiodicity estimator for high-quality speech
  synthesis,''
\newblock {\em Speech Communication}, vol. 84, pp. 57--65, 2016.

\bibitem{lancers}
``Lancers,'' \url{http://www.lancers.jp}.

\end{thebibliography}

\end{document}